\documentclass[conference]{IEEEtran}
\title{\large An Enhanced Spectral Efficiency Chaos-Based Symbolic Dynamics Transceiver Design}

\usepackage{flushend}
\usepackage{graphicx}
\usepackage{multicol}
\usepackage{amssymb}
\usepackage{epsfig}
\usepackage[cmex10]{amsmath}

\ifCLASSINFOpdf
\else
\fi
\begin{document}

%

\author{\IEEEauthorblockN{Georges Kaddoum$^{*}$,
Fran\c cois Gagnon$^{*}$ , Denis Couillard$^+$
}
\IEEEauthorblockA{ $^{*}$\'{E}cole de technologie sup\'{e}rieure, LaCIME Laboratory, Montreal, Canada }
\IEEEauthorblockA{$^+$ULTRA-TCS, Montreal, Canada\\
Email: georges.kaddoum[at]lacime.etsmtl.ca}
}

\maketitle

\footnotetext[1]{This work has been supported in part by  Ultra Electronics TCS and the Natural Science and Engineering Council of Canada as part of the 'High Performance Emergency and Tactical Wireless Communication Chair'  at \'{E}cole de technologie sup\'{e}rieure. (georges.kaddoum[at]lacime.etsmtl.ca)}

\begin{abstract}
\textbf{Chaotic synchronization performs poorly in noisy environments, with the main drawback being that the coherent receiver cannot be implemented in realistic communication channels. In this paper, we focus our study on a promising communication system based on chaotic symbolic dynamics. Such modulation shows a high synchronization quality, without the need for a complex chaotic synchronization mechanism. Our study mainly concerns an improvement of the bandwidth efficiency of the chaotic modulator. A new chaotic map is proposed to achieve this goal, and a receiver based on the maximum likelihood algorithm is designed to estimate the transmitted symbols. The performance of the proposed system is analyzed and discussed. }
\end{abstract}

\section{\textbf{Introduction}}
\label{sec:intro}

Using chaotic systems for wireless communications is one way to address some of the issues encountered when trying to avoid the detection and/or interception of signals. Chaotic signals are irregular, aperiodic, uncorrelated, broadband, and impossible to predict over longer time frames. These are properties that align well with some of the requirements for signals applied in communication systems for secure communications \cite{Ken00},\cite{Tam03}.

Generally, two different approaches can be used for chaotic signaling in digital communications. The first uses the real value of the chaotic signal to modulate the data information symbols \cite{Tam03,Kur05}. The second approach, proposed by Mazzini et al. \cite{Maz97}, quantizes the chaotic signal, which provides better performance than conventional spreading systems, but also leads to a loss of chaotic signal properties. To provide chaotic signal properties, the real value of the chaotic signal is used. Nonetheless, the general analysis which is carried out in this paper can easily be adapted for quantized chaotic signals.
A major challenge with coherent chaotic communication systems is the chaotic synchronization on the receiver side. The chaotic synchronization approach with two chaotic generators was proposed by Pecora and Carrol in \cite{Pec90}, and later used in coherent chaos-based communication systems in \cite{Cuo93} \cite{Cuo93a} \cite{Par92} \cite{Kol98} \cite{Kol97s} \cite{Lu02} \cite{Kol00a}. This approach involves coupling between the states of the two chaotic systems (coupled synchronization (CS)) to establish and maintain chaotic synchronization. Since the CS is highly sensitive to noise environments \cite{Kol98}, coherent communication systems using it on the receiver side show poor performances \cite{Kol98}. Further, over frequency-selective channels, the demodulation process is particularly difficult. Note that for the majority of coherent chaotic systems, two types of synchronization processes are needed to correctly accomplish demodulation, namely, the CS and phase synchronization. 

Non-coherent schemes, such as the differential chaos shift keying (DCSK) system proposed in \cite{Kol96} do not require chaotic synchronization on the receiver side. However, the security and the performance of the DCSK system are lower than those of coherent chaos-based communications systems. 
A promising chaotic signal modulation using symbolic dynamics was developed in \cite{Cif01} \cite{Lue05} and \cite{Sch012} for secure communication, which shows a high quality of synchronization. This type of chaotic modulation consists in mapping information bits to the state of a chaotic system through symbolic dynamics. Assigning information bits to states is not done arbitrarily. By dividing the phase space of the chaotic map into a finite number of partitions and assigning a symbol to each partition, the information bits are then embedded in the time evolution of the transmitted signal, with the partitions of the state space of the chaotic map being designated to represent the transmitted bits. For channel distortions, where intersymbol interference, as well as noise, are imposed on the transmitted sequence, optimal estimation and sequential channel equalization algorithms are developed in \cite{Cif01} to overcome these problems. In \cite{Sto97} and \cite{Dim00}, the synchronization process is analyzed and a high performance is achieved with symbolic dynamics. In our paper, we chose this type of chaotic modulation because of its robust and high quality of CS, which is demonstrated in the literature \cite{Sto97} \cite{Dim00}. 
On the receiver side, a maximum likelihood estimator based on a Viterbi decoding algorithm was introduced in \cite{Cif01} to estimate the chaotic sequence. Unfortunately, this approach uses a trellis with a large number of states. A simplified version of the trellis using a backward iteration of the system is presented in \cite{Lue05}. 

One challenge that remains when using chaotic systems is to improve bandwidth efficiency. The increasing demand for information and services has shown that limited bandwidth is a serious obstacle to the adoption of chaotic systems. The novelty of our paper lies in the fact that it increases the bandwidth efficiency of the system proposed in \cite{Lue05}. To that end, a new map is proposed, in which the symbols can be mapped onto a chaos-based multi-level modulator(i.e. M-QAM chaotic modulator) and a new receiver based on a Viterbi decoding algorithm is designed to decode the transmitted symbols. 

The paper is organized as follows: In Section II, the chaos-based M-QAM communication system with symbolic chaotic modulation and the receiver structure are explained, and the new proposed communication system is presented. The performance analysis of the proposed system is studied and a bit error rate expression is derived in section III. Section IV shows the simulation results. Finally, some remarks are given in the conclusion.

\section{\textbf{Chaos-based M-QAM communication system}}

This section presents the principle of symbolic dynamics; its application to chaotic modulation is explained, and finally, the proposed chaotic map within the receiver structure are detailed.

\subsection{\textbf{Symbolic dynamics}}

Symbolic dynamics was first used in digital communication in \cite{Hay93} \cite{Hao98}. By partitioning a chaotic phase space into arbitrary regions, and labelling each region with a specific symbol, the trajectories can be converted into a symbolic sequence. In this type of modulation, data information symbols are used to represent the trajectory of the chaotic map rather than generating a chaotic sequence directly by iteration (\ref{eq1}) to modulate the transmitted symbols:

\begin{equation} \label{eq1}
x[k] = f(x[k - 1]), \, \, \, \, x[k] \in I,
\end{equation}
\\
where $f()$ is a non-linear and non-invertible chaotic map and $I$ is the phase space. 

The state space $(I)$ of the chaotic map $f()$ is partitioned into $N$ disjoint regions, $I = \left\{ I \right\}_{i=1}^{N}$, such that $I_i \cap I_j = 0$ for $i \ne j$ and $\cup _{i = 1}^N I_i= I$. Note that this partition is not unique. For any sequence generated by iterating (\ref{eq1}), if we can assign $N$ alphabets $(\textbf{s} = [s_{1}; \:... ; \: s_{N}])$ to each of the disjoint regions, the dynamics of the system can be represented by a sequence of finite alphabet $\textbf{S}$. This sequence is called the symbolic dynamics of the system.

\begin{figure*}[htb!]
\centering
\includegraphics[width=14 cm]{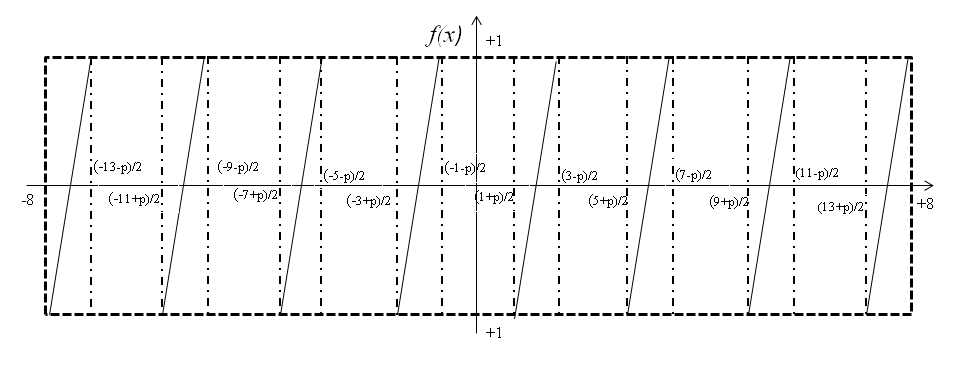}
\caption{Chaotic function $f(x)$}
\label{fig: c_func}
\end{figure*}

It was shown in \cite{Lue05} that the use of backward iteration onto the inverse function ($f^{-1}$) increases the quality of chaotic synchronization.
Because of the sensitivity of the chaotic map to the initial conditions, the sequence diverges rapidly, making the demodulation a real challenge. By iterating from a final condition $x[N]$ onto the inverse function ($f^{-1}$), the initial condition is contained in the set $ \cap _{n = 0}^{N - 1} f^{ - k} (I_i )$ \cite{Lue05}. When $N$ tends to infinity, the set contains a single initial condition, which shows a direct relation between the chaotic sequence and an infinite symbolic sequence. This is called backward iteration, and is well detailed in \cite{Sch011}. Under backward iteration, the chaotic map contracts, thus alleviating the problem of chaos synchronization since there is less sensitivity to initial conditions. From the framed transmitter, this is less of an issue since sampled sequences may be reversed. By using backward iteration, a long sequence will eventually converge toward an initial condition $x[0]$ (independent of the last sample $x[N]$) when guided by the sequence of symbols \cite{Lue05}. 
The symbolic sequence generated from backward iteration is denoted by:

\begin{equation}\label{eq4}
x[k] = f_{s[k]}^{ - 1} (x[k + 1]) = ... = f_{s[N - 1]}^{ - (N - k)} (x[N]),
\end{equation}
\\    
with $f_{s[k]}^{ - 1}$ being the inverse shift map.

\subsection{\textbf{Proposed chaotic modulator}}

In \cite{Lue05}, the spectral efficiency of the proposed chaotic modulator is low, with just a binary bit $b={+1, \: -1}$ capable of being coded. Our main goal is to improve this type of chaotic modulation because of its high security and quality of synchronization with low complexity. 

\begin{figure}[htb!]
\centering
\includegraphics[width=8 cm]{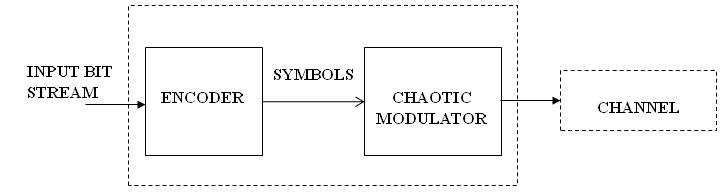}
\caption{Block diagram of a system for chaos-based M-QAM modulation at the transmitter end}
\label{M_QAM_s1}
\end{figure}

As shown in Figure \ref{M_QAM_s1}, to increase the data rate, we encode the input bit stream into M state symbols where $M=2^n$ is the amplitude level and $n$ is the number of bits per symbol. Finally, we map the symbols into a chaotic modulation to generate a chaos-based M-QAM signal. To achieve this goal, a chaotic map must be designed to accommodate the input symbols.

In our design, we study the case for $M=64$. The proposed chaotic map is an extension of the Bernoulli chaotic shift map. The corresponding piecewise linear chaotic map is given:

\begin{equation}\label{mf}
f(x) = \left\{ {\begin{array}{*{20}c}
   {\frac{{2x + 14}}{{1 - p}}\,\,\,for\,\,\,\frac{{ - 15 + p}}{2} \le x \le \frac{{ - 13 - p}}{2}}  \\
   {\frac{{2x + 10}}{{1 - p}}\,\,\,for\,\,\, \,  \,\, \frac{{ - 11 + p}}{2} \le x \le \frac{{ - 9 - p}}{2}}  \\
   {\frac{{2x + 6}}{{1 - p}}\,\,\,for\,\,\, \,\, \,\, \, \,\, \frac{{ - 7 + p}}{2} \le x \le \frac{{ - 5 - p}}{2}}  \\
   {\frac{{2x + 2}}{{1 - p}}\,\,\,for\,\,\, \,\, \,\, \,\,  \frac{{ - 3 + p}}{2} \le x \le \frac{{ - 1 - p}}{2}}  \\
   {\frac{{2x - 2}}{{1 - p}}\,\,\,  \,\, \,\,\, \, for\, \,\, \,\, \,\,\,\,\,\,  \frac{{1 + p}}{2} \le x \le \frac{{3 - p}}{2}}  \\
   {\frac{{2x - 6}}{{1 - p}}\,\,\,  \,\, \,\, for\,\,\, \,\, \,\,\,\,\,\, \,\, \frac{{5 + p}}{2} \le x \le \frac{{7 - p}}{2}}  \\
   {\frac{{2x - 10}}{{1 - p}}\,\,\,for\,\,  \,\, \,\, \,\, \,\, \,\,\,\frac{{9 + p}}{2} \le x \le \frac{{11 - p}}{2}}  \\
   {\frac{{2x - 14}}{{1 - p}}\,\,\,for\,\,\,   \,\,\,  \,\,\, \frac{{13 + p}}{2} \le x \le \frac{{15 - p}}{2}}  \\
\end{array}} \right.
\end{equation}  
\\
where $p$ is the control parameter of the map ($0 \le p < 1 $). The parameter $p$ controls the width between two consecutive regions of the map and the chaotic behavior of the generated sequences. Figure \ref{fig: c_func} shows the chaotic function of equation (\ref{mf}). The motivation for using this map resides in the adjustable control parameter $p$. When $p$ increases, the Euclidian distance between samples generated by the two consecutive intervals increases, which makes it possible to trade performance for security since it becomes easier to distinguish between samples of the two regions. The same map design is used to modulate the input symbols on the quadrature component (Q). With $8$ different space regions, this chaotic map is designed to obtain a chaos-based 64-QAM modulator.

As earlier mentioned, to provide for the backward iterations, we use the inverse function of equation (\ref{mf}):

\begin{equation}  \label{inv_qam}
f^{ - 1} (x) = \left\{ {\begin{array}{*{20}c}
   {\frac{{(1 - p)x - 14}}{2}\,\,\,\,for\,\,\,s = 0}  \\
   {\frac{{(1 - p)x - 10}}{2}\,\,\,\,for\,\,\,s = 1}  \\
   {\frac{{(1 - p)x - 6}}{2}\,\,\,\,for\,\,\,s = 2}  \\
   {\frac{{(1 - p)x - 2}}{2}\,\,\,\,for\,\,\,s = 3}  \\
   {\frac{{(1 - p)x + 2}}{2}\,\,\,\,for\,\,\,s = 4}  \\
   {\frac{{(1 - p)x + 6}}{2}\,\,\,\,for\,\,\,s = 5}  \\
   {\frac{{(1 - p)x + 10}}{2}\,\,\,\,for\,\,\,s = 6}  \\
   {\frac{{(1 - p)x + 14}}{2}\,\,\,\,for\,\,\,s = 7}  \\
\end{array}} \right.
\end{equation}

The improved chaotic modulator iterates backwards as per equation (\ref{inv_qam}) using the symbols sequence  $s_I[r]$ to define the iteration region $I[r]$ where $1 \le r \le 8$. In this paper, $ s_I[r]$ is the symbol sequence transmitted on branch $I[r]$. A similar sequence is constructed for branch $Q$ ($s_Q[r]$). According to the equation (\ref{inv_qam}), the generated chaotic signals can be seen in the $8$ regions on $I$. The inner
region between two consecutive regions is used as a guard region to ensure a minimum distance between the two waveforms associated with two successive symbols. 
The generated symbolic chaotic sequence at the output of the modulator on the branch $I$ is:
       
\begin{equation}\label{eq7}
 x[k] = f_{s_I[r]}^{ - 1} (x[k - 1])
\end{equation}

\begin{figure}[h!] 
\centering
\includegraphics[width=6 cm]{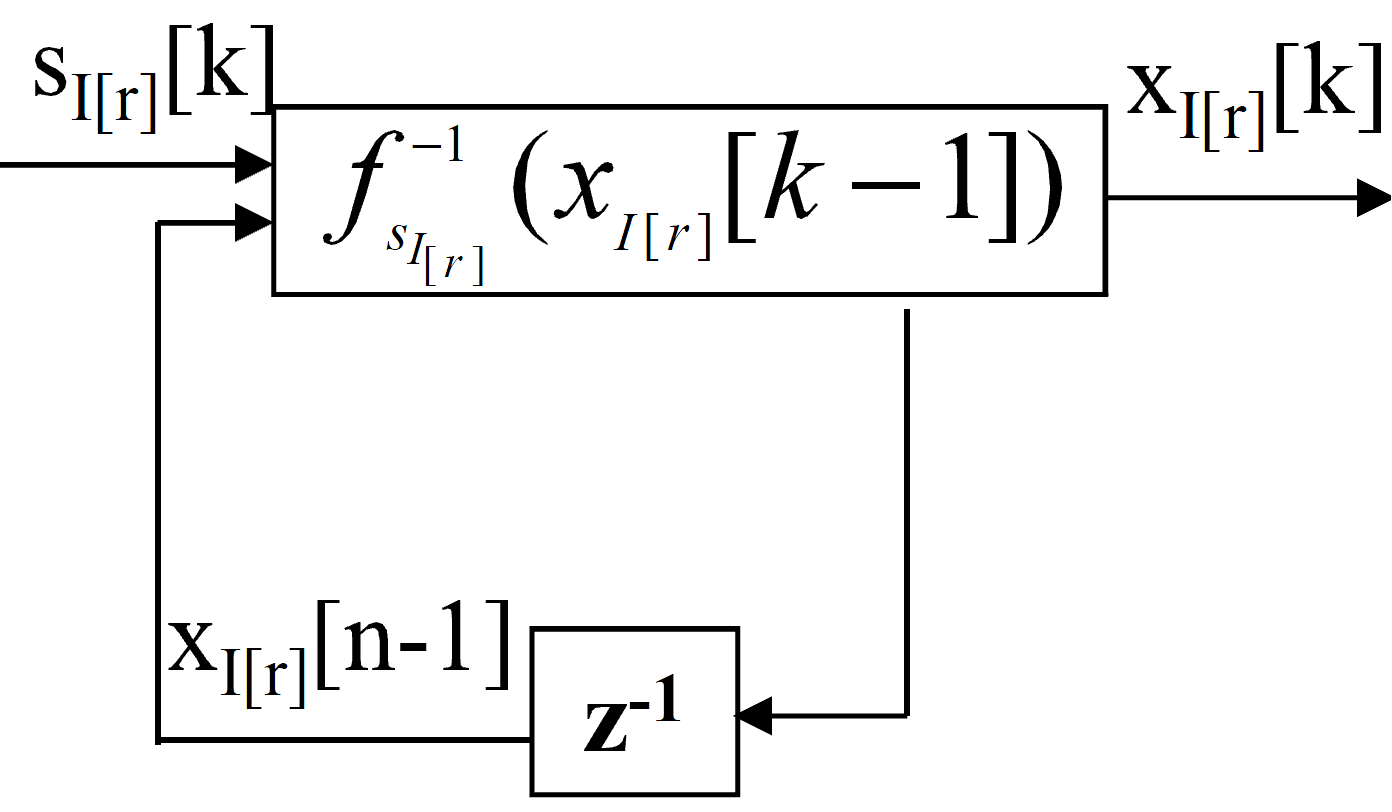}
\caption{Block diagrams illustrating the chaotic map sequences for an I input }
\label{M_QAM_si}
\end{figure}

Figure \ref{M_QAM_si} shows the M-QAM chaotic modulator. Note that the baseband chaotic signal at the output of the modulator can be moved to any desired frequency band for a passband transmission.

\begin{figure}[h!] 
\centering
\includegraphics[width=8 cm]{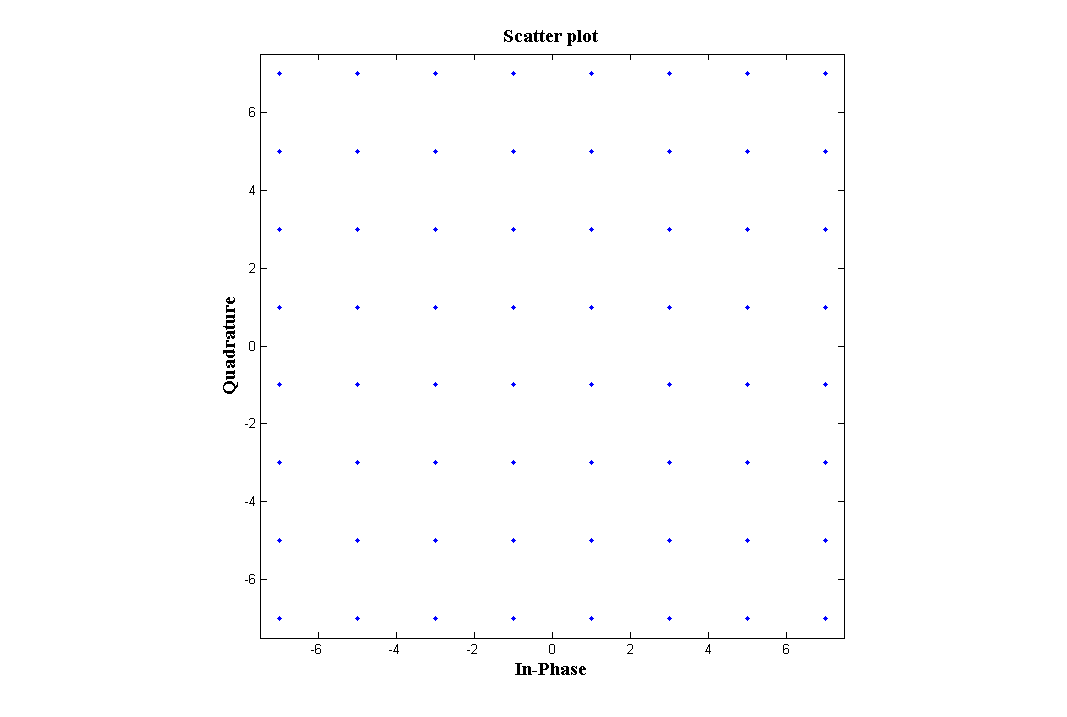}
\caption{Constellation of the 64-QAM chaotic modulator for p=1 }
\label{M_QAM_s2_1}
\end{figure}

\begin{figure}[h!]
\centering
\includegraphics[width=8 cm]{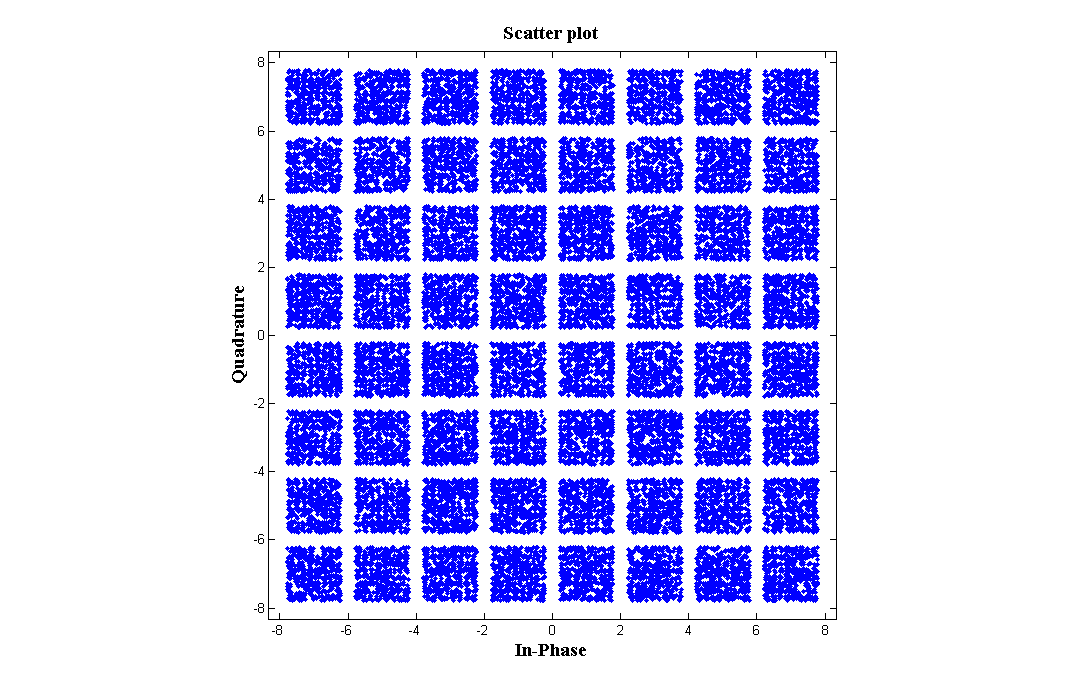}
\caption{Constellation of the 64-QAM chaotic modulator for p=0.8 }
\label{M_QAM_s2_2}
\end{figure}

\begin{figure}[h!] 
\centering
\includegraphics[width=8 cm]{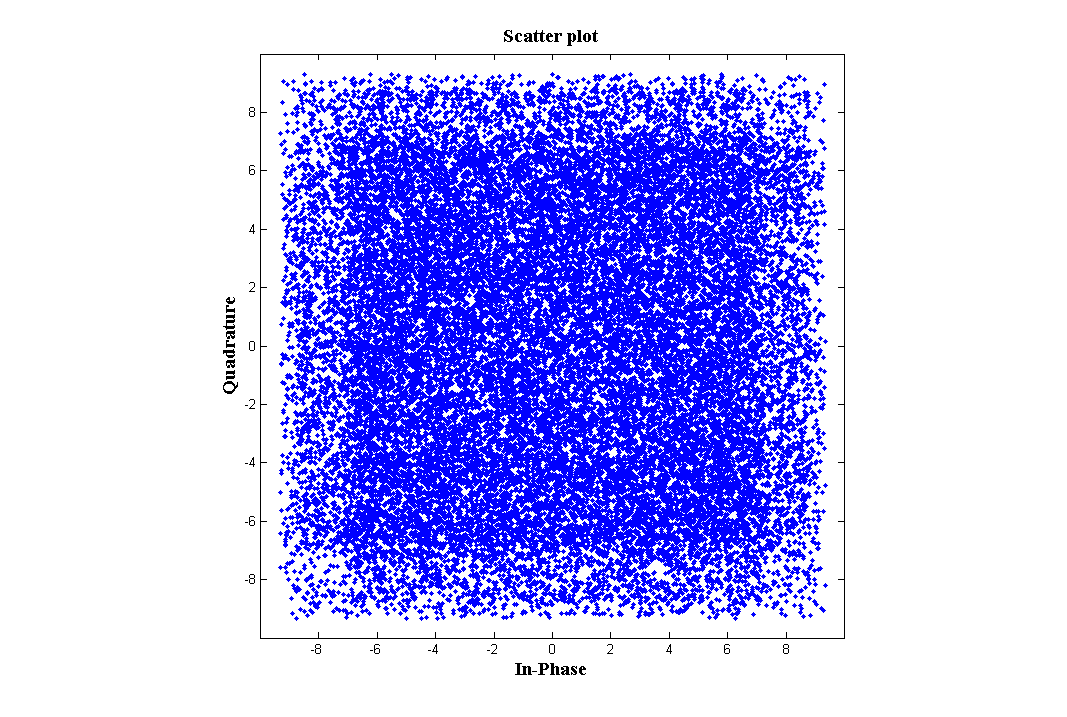}
\caption{Constellation of the 64-QAM chaotic modulator for p=0.4 }
\label{M_QAM_s2_3}
\end{figure}

Figures \ref{M_QAM_s2_1}, \ref{M_QAM_s2_2} and \ref{M_QAM_s2_3} show the different constellations of the chaos-based M-QAM modulator for different values of the control parameter $p$. It can clearly be seen that when $p$ tends to one, the constellation tends to the conventional 64-QAM, and when $p$ tends toward zero, the security of the transmitted symbols increases.

\subsection{\textbf{Receiver structure}}

A simplified Viterbi decoding algorithm with two states is fully discussed in \cite{Lue05}. In our case, we extend this algorithm to be operational with more than two states per iteration of chaotic sequence. 

After passing through an AWGN channel, the receiver signal is:

\begin{equation}\label{eq8}
 e[k] = x[k] +  w[k]
\end{equation}
\\
where  $ w[k]$  is an additive white Gaussian noise with power spectral density equal to $N_0/2$.

Generally, each state represents a possible symbolic sequence $S[m]$ of length $R$, and the branch metrics are: 

\begin{equation}
co_{ij} [k]=|e[k]-\Delta_{ij} [k]|^{2}
\end{equation}  
\\
where $co_{ij} [k]$ is the branch metric of taking the $j^{th}$ branch, starting from the $i^{th}$ node $(1\le i,j\le 8)$ at the $k^{th}$ time instant, and $\Delta_{ij} [k]$ is the sample resulting from the interference induced by $g(t)$, as:

\begin{equation}
{ \Delta}_{{ ij}} {  [k] = }\sum\limits_{{ m = 1}}^{{\rm R + 1}} {{ z}_{{ ij}} {  (m(T) )g(t - m(T + \varsigma) )}},
\end{equation}
\\
where  $g(t)$ is the  rectangular pulse shaping filter, $\varsigma$ represents the interference related to the sampling error,  $z_{ij} [m]=f_{(v[m])}^{(-1)} (x_i [m] )$ and $ v[m]=[\textbf{S}[m],j]$.


The cost of the $i$-th node at the $(k+1)$ time is computed from all the nodes at time $k$ as: 

\[
C_i [k + 1] = \mathop {\min }\limits_{j = 1,\,2..8} \left\{ {C_j [k] + c_{ji} [k]} \right\}.
\]

The Viterbi algorithm is then used to search through the $2^R$ states of the trellis for the most likely transmitted chaotic sequence. Finally, the estimated bits $\hat b [k]$ are obtained by selecting the best path of the transmitted sequence.

In our paper, the channel is an AWGN channel, and the sampling is perfect $\varsigma=0$. We use a filter without memory, $R=3$. In this case, the number of states is reduced to $8$.

\section{\textbf{Performance analysis }}

The optimal receiver for this type of transmission system is a trellis matched to the emitted code. The detector should estimate the correlation between the received signal and the expected signals, and select the sequence with the maximum total correlation according to the code constraint. This type of receiver is known as a Maximum Likelihood (ML) detection, and can be achieved in practice by applying the Viterbi algorithm. In our paper, we use ML detection to decode the received chaotic signal $e[k] = x[k] + w[k]$. The decoding process is based on the trellis diagram, where an eight-state Viterbi decoding algorithm is applied. For the equiprobable symbol $s$, the branch metrics are: 

\begin{equation}
c_{ij}[k]=\vert  r[k+1]- f_{j}^{-1}(\tilde x_{i}[k])  \vert^{2}
\end{equation}
\\
where $c_{ij}[k]$ is the cost of taking the $j-th$ branch, starting from the $i-th$ node $(1 \leq i, \, \, j \leq 8)$ at the time instant $(1\leq k \leq K)$, and $x_{i}[k]$ $(i \in {1,\,2... 8})$ is the sample obtained by iterating the function of equation (\ref{inv_qam}) backwards. The cost of the $i-th$ node at time $k+1$ is:

\begin{equation}
C_i [k + 1] = \mathop {\min }\limits_{j = 1,\,2...8} \left\{ {C_j [k] + c_{ji} [k]} \right\}
\end{equation}

The estimation method used for the received samples $\hat x[k]$  was developed in \cite{Sch011,Sch012}. The decoding algorithm maximizes the likelihood function $p(\hat x[k]\vert  x[k])$, which means that the probability that $\hat x[k]$  is estimated a priori, and which the assumption that $x[k]$ has been transmitted.

\subsection{Bit error rate expression}

The number of states of the Viterbi algorithm determines whether it is possible to carry out an analysis of the corresponding trellis in terms of error event distribution. The optimal receiver for this system is the soft Viterbi decoding algorithm, in which the Euclidean distance between the received signal and the corresponding signal of each state is computed. The computation results are used to feed a soft Viterbi decoder, which returns the ML sequence of the symbols according to the encoding chaotic map of the transmitter. 
To understand the performance analysis of our system, we compare the conventional M-QAM scheme with  our proposed system. To that end, we analyze the bit error performance of the Viterbi algorithm on the AWGN channel with a soft decision decoding of the M-QAM system.
We start by analyzing the simple case where the control parameter $p$ of the chaotic map tends toward $1$ ($p \rightarrow 1$). Here, the samples of the chaotic modulator of equation (\ref{inv_qam}) tend toward 64-QAM. The BER expression in this case is:

\begin{equation}\label{bpsk}
BER \approx 2\left( {1 - \frac{1}{{\sqrt M }}} \right)erfc\sqrt {\frac{{3nE_b }}{{2(M - 1)N0}}} 
\end{equation}
\\
where ${\frac{{E_b }}{{N_0 }}}$ is the bit energy-to-noise ratio, and $erfc$ is the complementary error function, and $n$ is the number of bits per symbol where $M=2^n$. In our case, $n=3$. 

Further, when the parameter $p$ is less than $1$ ($0\leq p <1$), the emitted bits are coded by representing each symbol with a chaotic sample associated with one of the eight regions of the chaotic map. Instead of having values of $+1$ or $-1$ when $p=1$ from the output of the chaotic modulator, in this case, we have a set of values which can take the values between the limits of each region. Since the transmitted bit energy is not constant after the coding by the chaotic modulator, we can consider that the equation (\ref{bpsk}) is the lower bound of our system. 

To derive an exact expression of the bit error rate, we analyze the effect of the chaotic modulator on the transmitted bit. The symbol is represented by a chaotic sample which can take a different number of values from the output of the chaotic modulator. These values of the chaotic samples lie within the eight intervals. Physically, this modulator can be seen as a fading variable which affects the transmitted bits. To establish a link between our system and the conventional 64-QAM system, we attribute the bit error rate expression over an AWGN channel:

\begin{equation}\label{bpsk_cod}
BER \approx 2\left( {1 - \frac{1}{{\sqrt M }}} \right)erfc\sqrt {\frac{{w_{\min }3n E_b }}{{2(M - 1)N0}}} 
\end{equation}
\\
where $w_{min}$ is a minimum code distance. 

The goal of the equation (\ref{bpsk_cod}) is to perform a general analysis of chaos communications systems based on symbolic dynamics modulation. In this case, $w_{min}$ in this model is the fading parameter.

Since the symbols are represented by chaotic samples, the Euclidean distance $d[k]=\sqrt{\vert (x[k])- f_{j}^{-1}(\tilde x[k])  \vert^{2}}$, can take a different value when the estimated symbol is different from that which is transmitted ($\hat s\left[ k \right] \ne s\left[ k \right]
$). The parameter $w_{min}$ is directly estimated from the variable $d[k]$.

For any given $K$ independent and identically distributed fading random variables with parameter $w_{min}$, the maximum likelihood estimate of $w_{min}$ is:

\begin{equation}\label{wmin}
\hat w_{\min }  = \sqrt {\frac{1}{{2K}}\sum\limits_{i = 1}^K {d[k]^2 } } 
\end{equation}

This estimation can be applied to any type of chaotic map. In our case, the chaotic map has a particular form, $\hat w_{\min }$ which can be computed directly  from the equation (\ref{inv_qam}).\\

The Euclidean distance is easily computed from (\ref{inv_qam}) of two successive regions is :

\[
\begin{array}{l}
 d = \sqrt {\frac{{(1 - p)x + (1 + p)}}{2} - \frac{{(1 - p)x - (1 + p)}}{2}}  \\ 
 d = \sqrt {(1 + p)}  \\ 
 \end{array}
\]
By replacing the Euclidean distance in  (\ref{wmin}), the expression of  parameter $\hat w_{\min }$ will be:

\begin{equation}
w_{\min }  \approx \sqrt {\frac{1}{2}(1 + p)} 
\end{equation}
It can clearly be seen that when the control parameter $p$  tends to $1$, the  parameter $\hat w_{\min }$ tend to one,  and the performance of the communications system tends to the performance of the 64-QAM system (i.e Figure \ref{perfs})

\section{\textbf{Simulation}}

\begin{figure}[h!]
\centering
\includegraphics[width=10 cm]{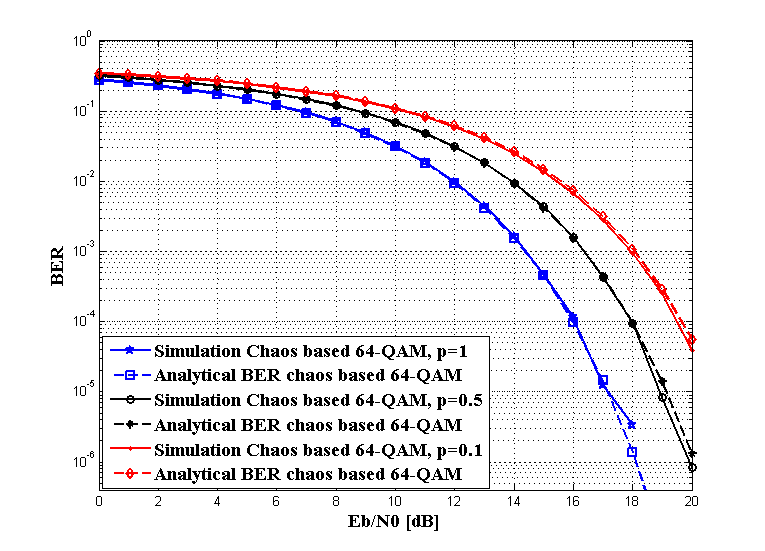}
\caption{BER performance of chaos-based M-QAM communication system for $p=0.1, \: 0.5 \: 1$  under AWGN channel}
 \label{perfs}
\end{figure}

In this section , we present different simulation results, in terms of BER, obtained with different values for the control parameter $p$, along with the theoretical results derived in the previous sections. To that end, Figure \ref{perfs} gives the theoretical and simulated BERs for $p=0.1, \: 0.5, \:1$. An excellent match can be seen the between analytical and simulated BERs, thus validating the analytical studies, along with the derivation of the corresponding parameter estimation $w_{min}$. Furthermore, simulation results prove that when $p$ tends to $1$, the performance of the system tends to the performance of the 64-QAM system. However, the motivation to use a chaotic code will have a number of structural advantages that may compensate for some performance degradation. Among other things, there is the advantage of extreme simplicity of generating spreading sequences and other aspects involving security.

\section{\textbf{Conclusion}}

In this paper, we focused our study on the improvement of the bandwidth efficiency of the chaotic modulator based on symbolic dynamics. To that end, a new chaotic map is proposed for achieving this goal. In our case, the chaotic function is designed to accommodate the input symbols and to obtain a chaos-based 64-QAM modulator. A receiver based on a maximum likelihood algorithm with 8 states is proposed to estimate the transmitted symbols. The performance of the proposed system is analyzed and discussed. A new methodology for computing the analytical BER expression in a mono-user transmission system based on chaotic symbolic dynamics is presented and analyzed. This method is based on an approximation of the minimum code distance $w_{min}$ by a fading parameter. The code distance $w_{min}$ is estimated from the Euclidian distance variable using the maximum likelihood estimator. For a mono-user system, there is an excellent match between the analytical and the simulated BERs, for all considered control parameters $p$. Simulation results show the accuracy of our approach. The performance of communication systems based on symbolic dynamics under an m-distributed fading channel is now under study.


\bibliographystyle{IEEEtran}
\bibliography{bibliographie}

\end{document}